\documentclass[a4paper]{article}
\usepackage[margin=3cm]{geometry}
\usepackage{latexsym}
\usepackage{amsmath}
\usepackage{amssymb}
\usepackage{amsfonts}
\usepackage[pdftex]{graphicx}

\newtheorem{theorem}{Theorem}
\newtheorem{remark}{Remark}
\newtheorem{definition}[theorem]{Definition}
\newtheorem{proposition}[theorem]{Proposition}

\begin{document}

\title{Chaotic behavior of a class of discontinuous dynamical systems of fractional-order}

\author{Marius-F. Danca\\
Dept. of Mathematics and Computer Science\\
              Avram Iancu University\\
              Str. Ilie Macelaru, nr. 1A, 400380, Cluj-Napoca,
              Romania\
              \and and \\
              Romanian Institute of Science and Technology,\\
              Str. Ciresilor nr. 29, 400487 Cluj-Napoca, Romania\\
              Tel.: +40-364-800171\\
              Fax: +40-364-800171\\
              danca@rist.ro
}

\maketitle

\begin{abstract} \indent In this paper the chaos persistence in a class of discontinuous
dynamical systems of fractional-order is analyzed. To that end, the
Initial Value Problem is first transformed, by using the Filippov
regularization \cite{Filippov}, into a set-valued problem of
fractional-order, then by Cellina's approximate selection theorem
\cite{Aubin si Cellina,Aubin si F}, the problem is approximated into
a single-valued fractional-order problem, which is numerically
solved by using a numerical scheme proposed by Diethelm, Ford \ and
Freed \cite{Diethelm}. Two typical examples of systems belonging to
this class are analyzed and simulated.
\end{abstract}

\textbf{Kkeyword} Fractional derivative; discontinuous dynamical
system; Filippov regularization; differential inclusion; numerical
method

\section{Introduction}

\indent Dynamical systems (d.s.), discontinuous with respect to the
state variable, occur in a large number of problems, especially from
mechanics (dry friction with stick and slip modes, impacts,
oscillating systems with viscous damping, elasto-plasticity,
alternatively forced vibrations, braking processes with locking
phases), electrical engineering (electrical circuits and networks
with switches, power electronics), automatic and optimal control
theory, convex optimizations, game theory, non-smooth control
systems synthesis, uncertain systems, walking machines, biological
and physiological systems and everywhere non-smooth characteristics
are used to represent switches (see e.g.
\cite{Andronov,Buhite,Clarke,Deimling,Schilling,Wiercigroch} and
some references therein). In other words, in the real word
non-smoothness is a common phenomenon. The underlying mathematical
models are generally described by a set of first-order differential
equations with discontinuous components. Particularly, the
discontinuity is due to the discontinuity of the state variable, of
the associated vector field, of the Jacobian, or to higher-order
discontinuity.

In practical examples, the discontinuity appears because of
piecewise continuous or switch-like functions (e.g. signum, absolute
value, Heaviside function -also known as the \textquotedblleft unit
step function\textquotedblright - maximum function, etc.).

On the other hand, there are many continuous d.s. which display
fractional-order dynamics. These mathematical phenomena allow to
describe a real object more accurately than do classical
\textquotedblleft integer\textquotedblright\ dynamics. One of the
main reasons to use the integer-order models was the absence of
methods for fractional differential equations.

The real objects or phenomena such as viscoelastic systems,
dielectric polarization, electrochemistry, percolation, material
science, polymer modeling, theory of ultra-slow processes,
electromagnetic waves, evolution of complex systems, financial
processes, secure communication, etc. are generally fractional (see
\cite{Bagley,Nakagava,Oustaloup,Sun,Podlubny,Ichise,Podlubny
2,Laskin,Kusnezov}).

We consider in this paper a class of unified Filippov discontinuous
d.s. of fractional-order and prove that the underlying Initial Value
Problem (I.V.P.) admits solutions which can be numerically
determined. We also prove numerically that in these systems of lower
than third order chaos may appear (as it is known, in the case of
integer order, according to the well known Poincaré-Bendixon
theorem, chaos appears only at systems of minimum order three).

These systems are modeled by the following autonomous I.V.P.
\begin{equation}
\frac{d^{q}x(t)}{dt^{q}}=g(x(t))+A~s(x(t)),~x(0)=x_{0},~t\in I=[0,\infty ),
\label{ivp generala}
\end{equation}
\noindent \smallskip with$~g:\mathbb{R}^{n}\rightarrow \mathbb{R}%
^{n},~A=\left( b_{i,j}\right) _{n\asymp n}$ real constant matrix,
the vector function $s:\mathbb{R}^{n}\rightarrow \mathbb{R}^{n}$
being composed of signum functions (the most encountered case), i.e.
\begin{equation}
s(x)=\left(
\begin{array}{c}
sgn(x_{1}) \\
\vdots \\
sgn(x_{n})%
\end{array}%
\right) ,  \label{s}
\end{equation}
\noindent $q$ the fractional order, $0<q\leq 1$ (although there are
also some applications for $q>1,~$ as is the case in many
applications).

The right-hand side depends on $p\in \mathbb{R},\mathbb{~}$the
control parameter$.$

Hereafter, we impose the following assumption on the discontinuous
right-hand functions

\noindent (\textbf{H1}):\label{H1}\emph{The right-hand side is
piecewise continuous (continuous on a finite number of }$m$\emph{\
open domains }$D_{i}\subset R^{n}$\emph{,}$~i=1,2,\ldots ,m$\emph{,
and has finite (possibly different)
limits from different boundary points, i.e. bounded discontinuities), }$g$%
\emph{\ being continuous (linear function in the great majority of practical
examples). The null set of the discontinuity (switch) points is }$%
M=R^{n}\backslash \cup D_{i}$\emph{.}

At every $M$ point, the discontinuity surface is a hyperplane (plane for $%
n=3).$

The considered class of systems is autonomous. Thus, the I.V.P.
(\ref{ivp generala}) can be written as follows
\begin{equation}
\frac{d^{q}x}{dt^{q}}=g(x)+A~s(x),~x(0)=x_{0},~t\in I.  \label{IVP}
\end{equation}
To achieve our goal, we consider the I.V.P. (\ref{IVP}) in a more
familiar framework, such as differential inclusions and differential
equations of fractional-order.

The paper is organized as follows: In Section 2, the discontinuous I.V.P. (%
\ref{IVP}) is transformed, via Filippov's regularization, into a
set-valued I.V.P. and then into a continuous differential equation
of fractional-order via Cellina's Theorem \cite{Aubin si
Cellina,Aubin si F}. In Section 3 the numerical integration
of differential equations of fractional-order, which will be used to integrate the obtained I.V.P. (\ref%
{IVP}), is presented briefly. In Section 4 this algorithm is used to
investigate the lowest-order $q$ for which chaos vanishes in two
representative chaotic systems.

\section{\label{continuous}Continuous approximation of I.V.P. (\protect\ref%
{IVP})}

\indent One of the first questions of set-valued analysis is how to
relate set-valued and single-valued functions, so as to avoid
dealing with the set-valued functions. Because the considered
examples in this paper are modeled by three-dimensional real I.V.P.,
the next results will be particularized in the Euclidean space
$\mathbb{R}^{n},$ some of them are valid even in more general (such
as Hilbert) spaces.

First, let us consider the IVP (\ref{IVP}) in the common case
$\emph{q}=1$ (i.e. discontinuous I.V.P. )
\begin{equation}
\frac{dx}{dt}=g(x)+A~s(x),~x(0)=x_{0},~t\in I.  \label{ivp cu q=1}
\end{equation}
Due to the discontinuity of the right-hand side, the I.V.P. may not
have any solution (see \cite{Filippov} for the background on
existence and uniqueness).

To overcome this situation, Filippov proposed the idea of restarting
the I.V.P. as a set-valued I.V.P. via differential inclusion
\cite{Filippov}
\begin{equation}
\frac{dx}{dt}\in F(x),~x(0)=x_{0},\text{ for a.a. }t\in I,  \label{dif incl}
\end{equation}
\noindent where $F:\mathbb{R}^{n}\rightrightarrows \mathbb{R}^{n}$
is a
convex set-valued vector function defined on the set of all subsets of $%
\mathbb{R}^{n}$ (for the background on set-valued functions we refer
e.g. to \cite{Aubin si Cellina,Aubin si F}).

Let us consider the following discontinuous I.V.P. enjoying the
assumptions \textbf{H1}
\begin{equation*}
\frac{dx}{dt}=f(x),\text{ }x(0)=x_{0},\text{ }t\in I.
\end{equation*}
\noindent One of the simplest definitions for $F$ is the following
\begin{equation}
F(x)=\bigcap\limits_{\varepsilon >0}\bigcap_{\mu
(M)=0}\overline{conv\left( f\{x\in \mathbb{R}^{n}:\parallel
x\parallel <\varepsilon \}\backslash M\right) },  \label{F}
\end{equation}
\noindent $F$ being the smallest closed and convex set containing
all limit values of $f$, and $~\mu ~$ the Lebesgue measure.

On $D_{i},~$where the function $f$ ~is
continuous, $F(x)$ consists of one point, which coincides with the value of $%
f$ at this point, $F(x)=\{f(x)\}$. Thus, on $D_{i}$, the I.V.P.
becomes a single-valued continuous problem $dx/dt=f(x).~$At the
discontinuity points,
the set $F(x)$ is a subset of $\mathbb{R}^{n}$ given e.g. by (\ref{F}).%

\begin{remark}
\label{remarka cu punctul} In the practical examples,$~\varepsilon $
must be small enough, so that the motion of the physical system can
be arbitrarily close to a certain solution of the differential
inclusion.
\end{remark}
For example, the set-valued version of the usual sign function is
defined by
\begin{equation*}
Sgn(x)=\left\{
\begin{array}{cc}
\{+1\} & x>0, \\
\left[ -1,1\right] & x=0, \\
\{-1\} & x<0.%
\end{array}%
\right.
\end{equation*}
By applying the Filippov regularization to the I.V.P. (\ref{ivp cu
q=1}), it becomes a set-valued I.V.P.
\begin{equation}
\frac{dx}{dt}\in g(x)+A~S(x),~x(0)=x_{0},~~for~ a.a.~t\in I,
\label{ste val pentru q=1}
\end{equation}
\noindent where $S$ is the set-valued form of $s~$(\ref{s}), i.e. $%
S(x)=\left( Sgn(x_{1}),Sgn(x_{2}),\ldots ,Sgn(x_{n})\right) ^{t}~$

\noindent Now, the obtained set-valued I.V.P. admits at least
a\emph{ generalized (Filippov) solution} (for existence and
uniqueness we refer to \cite{Filippov}
 and for numerical methods for differential inclusions \cite{Taubert,Dontchev
 Lempio}).
\begin{remark}
\label{usc}It is easy to see that the set-valued function $AS$ is upper
semicontinuous with closed and convex values (see e.g. \cite{Filippov} p. 85
or \cite{Aubin si Cellina} p. 101).
\end{remark}
Next, following the Filippov procedure applied to the right-hand
side of the I.V.P. (\ref{IVP}), one obtains the following
differential inclusion of fractional-order
\begin{figure}
\begin{center}
\includegraphics[clip,width=0.51\textwidth]{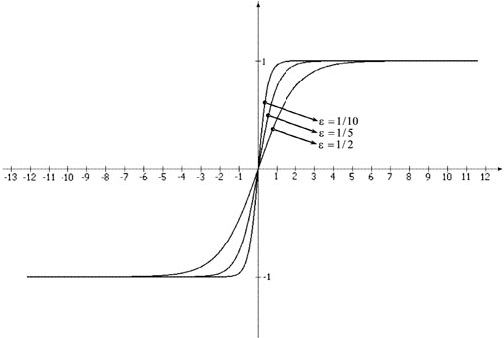}
\caption{Graph of the sigmoid function for different $\varepsilon$
values. }
\end{center}
\label{fig:1}       
\end{figure}

\begin{equation}
\frac{d^{q}x}{dt^{q}}\in g(x)+A~S(x),~x(0)=x_{0},~\text{for a.a.~}t\in I,
\label{aaaaa}
\end{equation}
\begin{definition}
\cite{Aubin si F},\cite{Kastner} \emph{selection }of a set-valued function $F:\mathbb{R}%
^{n}\rightrightarrows \mathbb{R}^{n}$ is a single-valued function $f:\mathbb{%
R}^{n}\longrightarrow \mathbb{R}^{n}$ satisfying\newline
\begin{equation*}
h(x)\in F(x),~~\forall x\in \mathbb{R}^{n}.
\end{equation*}
\end{definition}
Convex set-valued functions admit \emph{approximate selections }\cite%
{Aubin si F}\emph{,} i.e. single-valued functions $h_{\varepsilon }:\mathbb{R%
}^{n}\longrightarrow \mathbb{R}^{n}$ for which $Graph(h_{\varepsilon
})\subset Graph(F)+\varepsilon B~$and $h_{\varepsilon }(x)$ belongs to the
convex hull of the image of $F.~$
\begin{theorem}
\label{aprox}The set-valued function $AS~$ in the I.V.P. (\ref{aaaaa}%
)~admits a locally Lipschitzean approximate selection $h_{\varepsilon }:%
\mathbb{R}^{n}\longrightarrow \mathbb{R}^{n}$.
\end{theorem}

\noindent\textbf{Proof}
Because $AS$ is upper semicontinuous with closed convex values (Remark \ref%
{usc})$,$ according to Cellina's Theorem ( \cite{Aubin si F} p. 358,
T. 9.2.1) it admits an approximate locally Lipschitz selection.

The constructive proof of Cellina's theorem (see e.g. \cite{Aubin si
Cellina}) is a great advantage because it allows the construction of $%
h_{\varepsilon }~$.

There are several strategies to choose selections for a general
class of set-valued problem (see e.g. \cite{Kastner}).

Following the preceding theorem, we can enounce the following main
result
\begin{proposition}
\label{cont aprox}Let the I.V.P. (\ref{IVP}) with $g$ continuous. Then, the
I.V.P. can be approximated by the following continuous I.V.P. of
fractional-order\newline
\begin{equation*}
\frac{d^{q}x}{dt^{q}}=f(x), ~~x(0)=x_0, ~~t\in I,
\end{equation*}%
\newline
with $f:\mathbb{R}^{n}\longrightarrow \mathbb{R}^{n}$\newline
\begin{equation*}
f(x)=\left\{
\begin{array}{c}
g(x)+h_{\varepsilon }(x)~~\text{for~}x\in M\text{~,} \\
g(x)+As(x)~~\text{for~}x\notin M,%
\end{array}%
\right.
\end{equation*}%
\newline
where$~h_{\varepsilon }$ is an approximate selection of the
set-valued function $AS.$
\end{proposition}

\noindent\textbf{Proof}

The proof is obvious if we consider that first, the I.V.P.
(\ref{IVP}) is
transformed, via Filippov regularization, in the set-valued I.V.P. (\ref%
{aaaaa}). Next, Theorem \ref{aprox} can be applied in the
neighborhoods of the discontinuity points.

\begin{remark}
i) Obviously, $\varepsilon $ is a vector and in applications we have
chosen the same value for all its components;;\newline ii) As proved
in \cite{Danca 2}, this local approximation can be made even with
smooth functions;\newline iii) The proof of the above theorem could
also be done via maximal monotonicity of $AS$ which ensures its
closedness and convexity (see e.g. \cite{Aubin si Cellina} p.141,
Proposition 2).
\end{remark}

Because of the amount of necessary calculus for fractional numerical
integration, we shall use a simple form for $h_{\varepsilon }$.

To be precise, let us consider one component of $As$ in the I.V.P.
(\ref{IVP}),
the scalar function $s_{j}(x_{j})=a_{ij}sgn(x_{j}).$ In a $\varepsilon -$%
neighborhood of $x_{j}$,~ we have chosen in our numerical experiments for $%
h_{\varepsilon }$ the known \emph{sigmoid }function, widely used in
practical applications. Its scalar variant has the following form (
Fig. \ref{fig:1})
\begin{equation*}
h_{\varepsilon }\left( x\right) =\frac{2}{1+e^{-x/\varepsilon }}-1.
\end{equation*}
To connect continuously $h_{\varepsilon }~$in some $\varepsilon -$%
neighborhood of $x_{j},~$we shall use the following form
\begin{equation}
h_{\varepsilon }\left( x_{j}\right) =\frac{2a_{ij}}{1+e^{-x_{j}/\varepsilon }%
}-a_{ij}.  \label{sigmoid}
\end{equation}
The influence of $\varepsilon ~$is underlined in Fig. \ref{fig:2} d
and Fig. \ref{fig:2} e, where the visible (for $\varepsilon =1E-2$)
corners, typical for discontinuous d.s., become smooth if the width
of the neighborhood enlarges. Therefore, for physical reasons
$\varepsilon $ needs to be chosen relatively small.

\textbf{Conclusion} \emph{The discontinuity hyperplanes of I.V.P. (\ref{IVP}%
) can be replaced in their $\varepsilon$-neighborhoods with
continuous surfaces, via approximate selections}.

\section{Numerical approach of discontinuous systems of fractional-order}

\indent In this section we present summarily a numerical scheme for
continuous differential equations of fractional-order which will be
used for our I.V.P.(\ref{IVP}).
\subsection{\label{DDE}Numerical integration of differential
equations of fractional-order}

There are several definitions of the fractional differential
operator.

The \emph{Caputo fractional derivative} of order $q>0,$ where $q\in
(m-1,m]$ for some $m\in \mathbb{N}$ for a differential function
$x:[0,t]\longrightarrow \mathbb{R}$, is the most common tool, widely
used in real applications,
allowing for interpretation in a physically meaningful way (see e.g. \cite{Diethelm}, \cite%
{Podlubny} and \cite{Kai si Ford})
\begin{equation*}
D^{q}x=\frac{1}{\Gamma (m-q)}\int\limits_{0}^{t}\frac{x^{(m)}(s)}{%
(t-s)^{q+1-m}}ds,
\end{equation*}
\noindent where $\Gamma $ is the known Gamma function.

To be precise, let us consider in the sequel the continuous
autonomous fractional I.V.P.
\begin{equation}
\begin{array}{l}
D^{q}x=f(x),\text{ }0\leq t\leq T,\text{ }x^{(k)}(0)=x_{0}^{(k)}, \\
\text{ }k=0,1,\ldots ,m-1.%
\end{array}
\label{fract gene}
\end{equation}

\begin{center}
\begin{table*}[ht]
\caption{Pseudo-code of DFF numerical scheme} \label{tb:1} {\small
\hfill{} \mbox{}
\begin{tabular}{lll}
\hline $x_{i}[0]=x_{i0}[0],~i=1,2,\ldots ,N_{e}$ \ \ \ \ \ \ \ \ \
\{initial
conditions\} \\
for $\text{ }j:=1\text{ }$to$~N~$do \\
begin$\ \ \ \ $ \\
\ \ \ \ \ \ \ for$\text{ \ }i:=1~$to$~N_{e}~$do~\ \ \ \ \ \ \ \ \ \
\ \ \ \
\ \ \ \ \ \ $\ $\{predictor\} \\
$\ \ \ \ \ \ \ \ \ \ \ \ \ \ \ \ \
p_{i}:=~x_{i0}[0]+\dfrac{h^{q}}{\Gamma
(q+1)}\sum\limits_{k=0}^{j-1}b\left[ j-k\right] f_{i}\left( x_{1}\left[ k%
\right] ,\ldots ,x_{N_{e}}\left[ k\right] \right) ~~\ \ \ \ \ \ \ \ \ \ $ \\
\ \ \ \ \ \ \ for $i:=1$ to $N_{e}$ do $\ \ \ \ \ \ \ \ \ \ \ \ \ \
\ \ \ \
\ \ \ \ $\{corrector\} \\
$\ \ \ \ \ \ \ \ \ \ \ \ \ \ \ \ \ x_{i}\left[ j\right] :=x_{i~0}[0]+\dfrac{%
h^{q}}{\Gamma (q+2)}f_{i}(p_{1},\ldots ,p_{N_{e}})+(\left(
j-1\right) ^{q+1}-\ \left( j-1-q\right) j^{q})f\left( x_{1}\left[
0\right] ,\ldots
,x_{N_{e}}\left[ 0\right] \right) $ \\
\ \ \ \ \ \ \ \ \ $\ \ \ \ \ \ \ \ \ \ \ \ \ \ +\sum\limits_{k=1}^{j-1}a%
\left[ j-k\right] f\left( x_{1}\left[ k\right] ,\ldots ,x_{N_{e}}\left[ k%
\right] \right) $ \\
end \\
\hline
\end{tabular}}
\end{table*}
\end{center}
Under the continuity condition of the function $f$, there is at
least a solution (see e.g. \cite{Diethelm} where the most general
case of nonautonmous problems can be considered). Because $D^{q}$
has an $m$-dimensional
kernel ($m~$being just the value rounded up to the nearest integer i.e. $%
m=\lceil q\rceil )$, $m$ initial conditions need to be specified.
Certainly, for the usual case $0<q\leq 1$ we have to specify just
one condition.

The fractional differential operators are computationally time
expensive as compared to their integer-order counterparts. Some
numerical methods for (\ref{fract gene}), essentially ad hoc methods
for very specific types of differential equations are presented e.g.
in \cite{Podlubny}, \cite{Kai 2,Lubich,Shokooh}.

In agreement with the standard mathematical theory (\cite{Samko} Sect. 42),
the initial conditions for the I.V.P. (\ref{fract gene}) should have the form

\begin{equation*}
\frac{d^{q-k}}{dt^{q-k}}x\left( t\right)
|_{0_{+}}=b_{k},~~~k=1,\ldots ,m.
\end{equation*}

However, because in practical applications, these values are
frequently not available, and it may not even be clear what their
physical meaning is, one can specify the initial conditions in the
classical form as they are commonly used in initial value problems
with integer-order equations \cite{Kai si Ford}.

The numerical scheme to integrate (\ref {fract gene}), which is used
in this paper, is a predictor-corrector algorithm (called hereafter
\emph{DFF}), proposed by Diethlem, Ford and Freed in \cite{Diethelm}
together with error estimates and numerical examples. The scheme is
a generalization of the classical multistep method
Adams--Bashforth--Moulton and uses the Caputo derivative.

Being of practical significance and helpful for solving a broad
class of problems, the method has been constructed and analyzed for
the fully general set of equations without any special assumptions,
and is easy to implement on a computer

Let us next assume that we are working on a uniform grid $%
\{t_{n}=nh:n=0,1,...,N\}$ with some integer $N$ with the step size
$h:=T/N$ and the case of $q\in (0,1].$

The predictor formula for $\ $the value at the point $t_{n+1},x^{p},$ is the
fractional variant of the Adams-Bashforth method which for our particular case $%
m=1$, has the expression
\begin{equation*}
x^{p}\left( t_{n+1}\right) =x_{0}+\frac{1}{\Gamma \left( q\right) }%
\sum\limits_{j=0}^{n}b_{j,n+1}f\left( x\left( t_{j}\right) \right) ,
\end{equation*}
\noindent while the corrector formula (the fractional variant of the
one-step implicit Adams Moulton method)
\begin{eqnarray*}
x\left( t_{n+1}\right) &=&x_{0}+\frac{h^{q}}{\Gamma (q+2)}f\left(
x^{p}\left( t_{n+1}\right) \right) \\
&&+\frac{h^{q}}{\Gamma (q+2)}\sum\limits_{j=0}^{n}a_{j,n+1}f\left( x\left(
t_{j}\right) \right) ,
\end{eqnarray*}
\noindent where $\Gamma $ is the Gamma function, and $a$ and $b$ are
the corrector and predictor weights respectively given by the
following formula

\begin{equation}
a_{j,n+1}=\left\{
\begin{array}{ll}
n^{q+1}-\left( n-q\right) \left( n+1\right) ^{q} & \text{if }j=0, \\
\left( n-j+2\right) ^{q+1}+\left( n-j\right) ^{q+1}\\ -2\left(
n-j+1\right)
^{q+1} & \text{if }1\leq j\leq n, \\
1 & \text{if}~j=n+1,%
\end{array}%
\right.  \label{a}
\end{equation}
\noindent and
\begin{equation}
b_{j,n+1}=\frac{h^{q}}{q}\left( \left( n+1-j\right) ^{q}-\left( n-j\right)
^{q}\right) .  \label{b}
\end{equation}
After a few modifications, designed to enhance the efficiency
\cite{Diethelm}, \emph{DFF} scheme has the form presented as
pseudo-code in Table \ref{tb:1} for the case $m=1~$ and a set of
$N_{e}~$autonomous differential equations of fractional-order $q$
with the right-hand side

 $%
\left( f_{1}\left( x_{1},x_{2},\ldots ,x_{N_{e}}\right)
,\ldots,\text{~}f_{N_{e}}\left( x_{1},x_{2},\ldots ,x_{N_{e}}\right)
\right) ^{t}$

\noindent where:

$x_{i0},~i=1,2,...,N_{e}$ are the initial conditions;

$x[\cdot ]$ array of $N+1$ real numbers containing the approximate
solutions: $x_{k}$ $\thickapprox x(kh),$ $x$ being the exact
solution;

$p$ the predicted values;

$a,b,$ arrays of $3\times (N+1)$ real numbers which, in this
algorithm, are determined with the following variant
\cite{Diethelm}.

\begin{table}[h]
\begin{center}
\begin{tabular}{lll}
\hline

for~$k:=1~$to$~N~$do \\
$\ \ \ \ a[k]:=k^{q}-\left( k-1\right) ^{q}$ \\
$\ \ \ \ b[k]:=\left( k+1\right) ^{q+1}-2k^{k+1}+\left( k-1\right) ^{q+1}$
\\
end \\
\hline

\end{tabular}%
\end{center}
\end{table}

The Gamma function $\Gamma ~$can be approximated via the so-called
Lanczos approximation \cite{C} using the formula
(http://www.rskey.org/gamma.htm)
\begin{equation}
\Gamma (z)=\frac{\sum\limits_{i=0}^{6}P_{i}z^{i}}{\prod\limits_{i=0}^{6}(z+i)%
}(z+5.5)^{z+0.5}e^{-(z+5.5)},  \label{gama}
\end{equation}
\noindent for a complex variable $z$ for which $Re(z)>0$. The
coefficients $P$ are drawn in Table \ref{tab:3}
\begin{table}[h]
\begin{center}
\caption { $P$ coefficients for $\Gamma$ function  (\ref{gama})}
\label{tab:3}
\begin{tabular}{l}
\hline\noalign{\smallskip}
$P_{0}=75122.6331530$ \\
$P_{1}=80916.6278952$ \\
$P_{2}=36308.2951477$ \\
$P_{3}=8687.24529705$ \\
$P_{4}=1168.92649479$ \\
$P_{5}=83.8676043424$ \\
$P_{6}=2.50662827511$ \\
 \noalign{\smallskip}\hline
\end{tabular}
\end{center}
\end{table}
\begin{remark}
i) Another way to deal with fractional derivatives is the use of the
frequency domain approximation, based on the approximation of the
fractional operators in the frequency domain. This technique,
proposed by researchers on automatic control, is suitable for
Simulink (Matlab) (see e.g.\cite{Charef,Ivo,Wajdi});\newline ii) As
known, an attractor requires a relatively large integration time
interval (for example for some of our experiments one needs
$N=2E4\div 2.5E4$), and an array of this size is hard to implement
in any compilers. The best solution we found for our examples was to
save $x$ into a file.\newline
\end{remark}
\subsection{Numerical integration of I.V.P. (\protect\ref{IVP})}
Now, using the results in Section 2 and Subsection 3.1, we can
enounce the main result of this paper
\begin{theorem}
Let I.V.P. (\ref{IVP}) with $g$ continuous. Then the problem can be
numerically integrated.
\end{theorem}

\noindent\textbf{Proof}

The I.V.P. (\ref{IVP}) can be transformed into a set-valued I.V.P.,
via Filippovþs regularization (Section \ref{continuous}) which can
be continuously approximated (Proposition \ref{cont aprox}). The
obtained
single-valued I.V.P. may be numerically integrated via \emph{DFF }(Section %
\ref{DDE}).\newline

One of the most unpleasant aspects of the DFF scheme is that it
requires each step, the entire backward integration history. In
other words, each value $x_{k}$ is a determined function of all
previous values $x_{0},$ $x_{1},\ldots ,x_{k-1}$

\begin{figure*}[!ht]
 \begin{center}
\includegraphics[clip,width=0.98\textwidth]{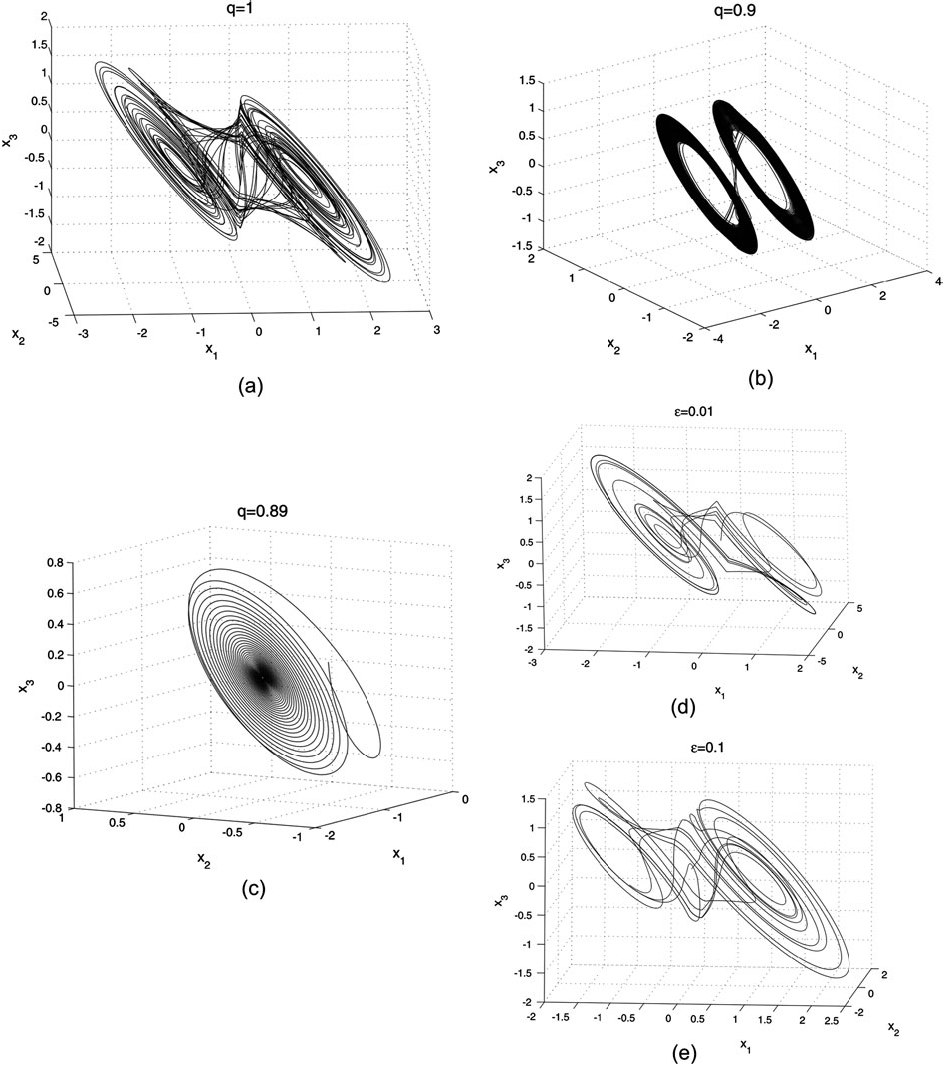}
\caption{Phase plots for fractional discontinuous Sprott system (\ref{Sprott}%
). a),b) Chaotic trajectories for $q=1$ and $q=0.9.$ c) Attractive
fixed point for $q=0.89.$ d) A chaotic trajectory for $q=0.95$ and
$\varepsilon =0.01$ e) A chaotic trajectory for $q=0.95$ and
$\varepsilon =0.1$.}
\label{fig:2}       
\end{center}
\end{figure*}

or, for a large $k$, this implies supplementary computational cost.
Thus, unlike the classical derivatives of integer order, the
fractional derivatives operators are not local (i.e. we cannot
determine $D^{q}x(t)$ by using only a few values of $x$ in a
neighborhood of $t$).
\begin{remark}
 Lipschitz continuity of $g$ is necessary only to ensure the
uniqueness either of the I.V.P. (\ref{ste val pentru q=1} ) or
(\ref{IVP}) if the study of convergence properties and errors of
some numerical method is required.
\end{remark}
Error analysis is a difficult task, especially because of the
numerical integration of fractional equations (see a detailed study
in \cite{Kai si Ford}). Therefore, in our study we have chosen
empirically $h$ so that, for the common case $q=1$, the obtained
attractor fits as well possible the known attractor shape in the
phase space $\mathbb{R}^{3}$. Also, the optimal choice of the step
length must ensure the maximum accuracy in the approximate solution
at minimum computational cost.

In order to avoid the specific phenomena for discontinuous d.s. in
the discontinuity surfaces (e.g. sliding modes \cite{Wiercigroch}),
the initial conditions need to be chosen in $D_{i}.~$As can be seen
from the obtained images, the trajectories present some "corners"
typical of discontinuous d.s. \newline

\section{Examples}

\indent Let us first introduce the following notion

\textbf{Notation} Let us denote by $d=3q$ the \emph{order} of the
d.s. modeled by I.V.P. \ref {IVP} and by $d^{*}$ the lowest value
for which chaos persists.

 While
in the case of the continuous d.s. of integer-order chaos can appear
only in nonlinear systems with order minimum 3, in nonlinear systems
of fractional-order, this is not the case. For example, in
\cite{Hartley} has been shown that a Chua circuit of order $d$ as
low as 2.7 can produce a chaotic attractor; in \cite{Arena} it is
shown that a sinusoidally nonautonomous driven Duffing system with
$d$ less than 2 can still behave in a chaotic manner; in
\cite{Xiang} it is proved that the lowest order $d$ for fractional
Lorenz system is $d^{*}=2.97$ and in \cite{Jun} it is proved that
the lowest-order chaotic system out of all the found chaotic systems
reported in specialized literature, appears in the fractional L\"{u}
system.

In this section we explore the chaos disappearance, by searching
$d^{*}$ with the algorithm proposed before, in two examples, of
known discontinuous three-dimensional d.s. In the simulations, we
only vary the fractional derivative order $q$ while the control
parameter $p$ is fixed so the underlying d.s. of integer order
behaves chaotically. Simulations are performed by lowering $q$ from
$q=1~$(the integer case) until chaos behavior disappears.

All computations in this paper were performed in double precision
arithmetic, $N$ was chosen large enough (of order $10E4$) to provide
accurate
results, the integration step size $h$ of $1E-3~$and $%
\varepsilon =1E-3$ of order.
\begin{remark}
In all the practical examples we have found in the scientific texts
\ (either two-dimensional, typical for dry friction problems, or
three-dimensional, found generally in many areas of science) there
is at most one $signum$ function in each equation of I.V.P. This
helps to facilitate the numerical integration (as well known, it is
time-consuming to compute the exponential functions).
\end{remark}

\begin{figure*}
\includegraphics[clip,width=1\textwidth]{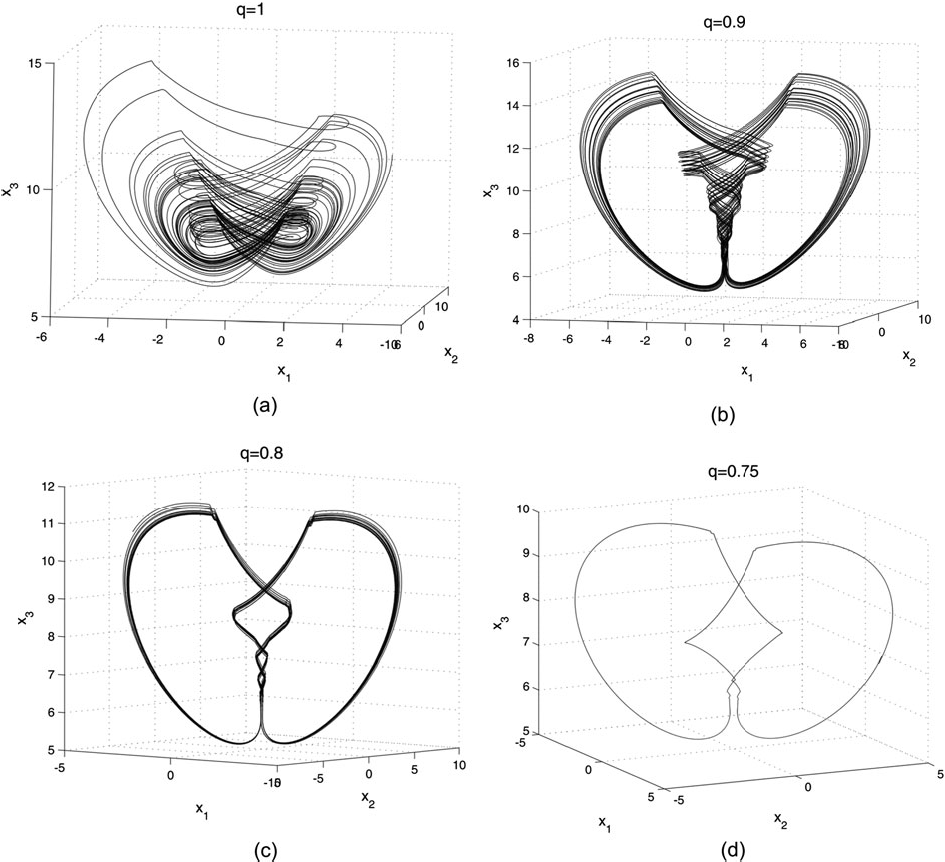}
\caption{Phase plots for fractional discontinuous Chen system
(\ref{Chen}). a)-c) Chaotic trajectories for $q=1,~q=0.9$
and$~q=0.8$ respectively. d) Stable limit cycle for $q=0.75.$ }
\label{fig:3}       
\end{figure*}

\subsection{Sprott system}

\indent The first example considered in this paper was proposed by
Sprott in  \cite{Wajdi} and \cite{Sprott} and it is a representative
one for the considered class of d.s.
\begin{eqnarray}
\frac{d^{q}x_{1}}{dt^{q}} &=&x_{2},  \label{Sprott} \\
\frac{d^{q}x_{2}}{dt^{q}} &=&x_{3},  \notag \\
\frac{d^{q}x_{3}}{dt^{q}} &=&-x_{1}-x_{2}-px_{3}+sgn(x_{1}).  \notag
\end{eqnarray}
whith
\begin{equation*}
g(x)=\left(
\begin{array}{c}
x_{2} \\
x_{3} \\
-x_{1}-x_{2}-px_{3}%
\end{array}%
\right) ,\text{ }A=\left(
\begin{array}{ccc}
0 & 0 & 0 \\
0 & 0 & 0 \\
1 & 0 & 0%
\end{array}%
\right) .\text{ }
\end{equation*}

In this case, the chaos corresponding to $p=0.5$ (see Fig. \ref{fig:2} a for $q=1$ and Fig. \ref {fig:2} b for $q=0.9$) vanishes at $%
q=0.89~$ when $d^{*}=2.57$ and when an attractive fixed point is
obtained (Fig. \ref{fig:2} c).

\subsection{Chen system}
\indent The next example is the fractional variant of the
discontinuous Chen system \cite{Chen} belonging to a more general
class of discontinuous fractional systems
\begin{eqnarray}
\frac{d^{q}x_{1}}{dt^{q}} &=&-p(x_{2}-x_{1}),  \label{Chen} \\
\frac{d^{q}x_{2}}{dt^{q}} &=&sgn(x_{1})(7-p-x_{3})+0.7x_{2},  \notag \\
\frac{d^{q}x_{3}}{dt^{q}} &=&sgn(x_{2})x_{1}-0.168x_{3}.  \notag
\end{eqnarray}

For $q=1$ the attractor (chaotic for $p=1.18$) is presented in Fig.
\ref{fig:3} a. In this chase chaos seems to be more "persistent"
(see Fig. \ref{fig:3} b,c) since it exists until $q=0.75$ (see Fig.
\ref{fig:3} d) when a stable limit cycle appears. Thus,
$d^{*}=2.25$.

\textbf{Conclusion} In this paper, a continuous approximation, via
Filippov regularization, of a class of fractional-order
discontinuous I.V.P. has been considered. The obtained I.V.P. of
fractional-order was integrated by using the algorithm proposed by
Diethlem, Ford and Freed in \cite{Diethelm}. Discontinuous
fractional-order Sprott and Chen systems have been considered. We
have found that chaos still exists in these systems if the order is
lower than three. In the cases of continuous fractional systems, the
chaotic behavior is affected when the fractional-order decreases.
However, in our opinion, choosing the optimal value for $q$ (and
implicitly for $d$), so that the mathematical model (\ref {IVP}) can
better illustrate the behavior and properties of a physical system,
is much more important than finding the lowest $d^{*}$ (a simple
reason could be that for $d=d^{*}$ chaos seems to vanish, in
possible contradiction with the real system).


\begin{thebibliography}{}
\bibitem{Filippov} Filippov, A.F.: Differential Equations with Discontinuous
Right-Hand Sides. Kluwer Academic Publishers, Dordrecht (1988)

\bibitem{Aubin si Cellina} Aubin, J.-P., Cellina, A.: Differential inclusions
set-valued maps and viability theory. Springer, Berlin (1984)

\bibitem{Aubin si F} Aubin, J.-P., Frankowska, H.: Set-valued analysis.
Birkh$\ddot{a}$user, Boston (1990)

\bibitem{Diethelm} Diethelm, K., Ford, N.J. Freed, A.D.: Predictor-Corrector
Approach for the Numerical Solution of Fractional Differential
Equations. Nonlinear Dynamics \textbf{29}, 3-22 (2002)

\bibitem{Andronov} Andronov, A.A, Vitt A.A, Khaikin, S.E.: Theory of oscillators.
Pergamon Press, Oxford (1966)

\bibitem{Buhite} Buhite, J.L., Owen, D.R.: An ordinary differential equation from
the theory of plasticity. Arch. Ration. Mech. Anal. \textbf{71},
357-383 (1979)

\bibitem{Clarke} Clarke, F.H.: Optimization and non-smooth analysis.
Wiley, New York (1983)

\bibitem{Deimling} Deimling K.: Multivalued differential equations and dry friction
problems. In: Fink, A.M., Miller, R.K., Kliemann W., editors. Proc.
conf. Delay and Differential Equations, pp. 99-106. World
Scientific, Singapore (1992)

\bibitem{Schilling} Schilling, K.: An algorithm to solve boundary value
problem for differential equations and applications in optimal
control. Numer. Funct. Anal. Optim. \textbf{10}, 733-764 (1989)

\bibitem{Wiercigroch} Wiercigroch, M:, de Kraker B. Applied nonlinear dynamics
and chaos of mechanical systems with discontinuities. World
Scientific, Singapore (2000)

\bibitem{Bagley} Bagley, R.L., Calico, R.A.: Fractional order state equations for
the control of viscoelastically damped structures. J. Guid. Control
Dynam. \textbf{14}, 304-311 (1991)

\bibitem{Nakagava} Nakagava, M., Sorimachi, K.: Basic characteristics of a
fractance device. IEICE T. Fund. Electr. E75-A(12), 1814-1818 (1992)

\bibitem{Oustaloup} Oustaloup, A.: La Derivation Non Entiere: Theorie,
Synthese et Applications. Hermes, Paris (1995)

\bibitem{Sun} Sun, H.H., Abdelwahad, A.A., Onaral, B.: Linear approximation of
transfer function with a pole of fractional order. IEEE Trans.
Automat. Contr. \textbf{29},441-444 (1984)

\bibitem{Podlubny} Podlubny, I.: Fractional Differential Equations. Academic
Press, San Diego (1999)

\bibitem{Ichise} Ichise, M., Nagayanagi Y., Kojima, T.: An analog simulation of
noninteger order transfer functions for analysis of electrode
process. J. Electroanal Chem, \textbf{33}, 253-265 (1971)

\bibitem{Podlubny 2} Podlubny, I., Petr\'{a}\v{s}, I., Vinagre, B.M.,
O'Leary, P., Dorc\'{a}k, L'.: Analogue realization of
fractional-order controllers. Nonlinear Dyn. \textbf{29}(1-4),
281-296 (2002)

\bibitem{Laskin} Laskin, N.: Fractional market dynamics. Physica A
\textbf{287}, 482-492 (2000)

\bibitem{Kusnezov} Kusnezov, D., Bulgac, A., Dang, G.D.: Quantum Levy processes and
fractional kinetics. Phys. Rev. Lett. \textbf{82}, 1136-1139 (1999)


\bibitem{Taubert} Taubert, K.: Converging multistep Methods for Initial Value
Problems Involving Multivalued Maps. Computing \textbf{27}, 123-136
(1981)

\bibitem{Dontchev Lempio} Dontchev, A. and Lempio, F.: Difference Methods
for Differentiale Inclusions. SIAM Rev., \textbf{34}, 263-294 (1992)

\bibitem{Kastner} Kastner-Maresch A., Lempio F.: Difference methods with
selection strategies for differential inclusions. Numer. Funct.
Anal. Optimiz. \textbf{14}, 555-572 (1993)


\bibitem{Danca 2} Danca M.-F., Codreanu, S.: On a possible approximation of
discontinuous dynamical systems. Chaos, Solitons \& Fractals
\textbf{13}, 681-691 (2002)


\bibitem{Kai si Ford} Diethelm, K., Ford, N.J.: Analysis of Fractional
Differential Equations. Journal of Mathematical Analysis and
Applications \textbf{265}, 229-248 (2002)

\bibitem{Kai 2} Diethelm, K.: An algorithm for the numerical solution of
differential equations of fractional order. Electronic Transactions
on Numerical Analysis \textbf{5}, 1-6 (1997)

\bibitem{Lubich} Lubich, C.: Discretized fractional calculus. SIAM Journal
on Mathematical Analysis \textbf{17}, 704-719 (1986)

\bibitem{Shokooh} Shokooh, A., Suarez, L.E.: A comparison of numerical
methods applied to a fractional model of damping materials. J. Vib.
Control, \textbf{5}, 331-354 (1999)

\bibitem{Samko} Samko, S. G., Kilbas, A. A., Marichev, O. I.:
Fractional Integrals and Derivatives: Theory and Applications.
Gordon and Breach, Yverdon, (1993)


\bibitem{C} Press, W.H., Teukolsky, S.A., Vetterling, W.T.,
Flannery, B.P.: Numerical Recipes in C, The Art of Scientific
Computing, Second Edition. Cambridge University Press, (1992)

\bibitem{Charef} Charef, A. Sun, H.H., Tsao, Y.Y. Onaral, B.: Fractal system as represented by singularity function. IEEE T. Automat. Contr. \textbf{37}, 1465-1470 (1992)

\bibitem{Ivo} Petr\'{a}\v{s}, I.: Chaos in the fractional-order Volta's
system: modeling and simulation. Nonlinear Dynam. 57(1-2), 157-170
(2009)

\bibitem{Wajdi} Ahmad, W.M., Sprott, J.C.: Chaos in
fractional-order autonomous nonlinear systems. Chaos, Solitons and
Fractals \textbf{16}, 339-351 (2003)


\bibitem{Hartley} Hartley, T.T., Lorenzo, C.F., Qammer, H.K.: Chaos in a fractional
order Chua's system. IEEE T. Circuits-I \textbf{42}(8), 485-490
(1995)

\bibitem{Arena} Arena, P., Caponetto, R., Fortuna, L., Porto, D.: Chaos in a
fractional order Duffing system. In: Proceedings ECCTD, Budapest,
September, 1259-1262 (1997)

\bibitem{Xiang} Wu, X.-J., Shen, S.-L.: Chaos in the
fractional-order Lorenz system. International Journal of Computer
Mathematics \textbf{86}(7), 1274-1282 (2009)

\bibitem{Jun} Lu, J.G.: Chaotic dynamics of the fractional-order L\"{u}
system and its synchronization. Physics Letters A \textbf{354},
305-311 (2006)

\bibitem{Chen} Aziz-Alaoui, M.A., Chen, G.: Asymptotic Analysis of a New
Piecewise-Linear Chaotic System. Int. J. Bifurcat. Chaos
\textbf{12}(1), 147-157 (2002)

\bibitem{Sprott} Sprott, J.C.: A New Class of Chaotic Circuit. Physics
Letters A, \textbf{266}, 19-23 (2000)

\end{thebibliography}
\end{document}